\documentclass[aps,pre,nofootinbib,twocolumn,superscriptaddress,
               amsmath,amssymb,floatfix,showpacs]{revtex4}

\usepackage{amsmath} \usepackage{epsfig} \usepackage{bm}

\def\bc{\begin{center}}
\def\ec{\end{center}}
\def\be{\begin{equation}}
\def\ee{\end{equation}}

\def\bear{\begin{eqnarray}}
\def\eear{\end{eqnarray}}

\newcommand{\C}{\mathcal{C}}

\newcommand{\phibar}{\overline{\phi}}

\begin{document}

\title{Dynamic first-order phase transition in kinetically constrained
models of glasses}

\author{J.P. Garrahan}

\affiliation{School of Physics and Astronomy, University of
Nottingham, Nottingham, NG7 2RD, UK}

\author{R.L. Jack}

\affiliation{Department of Chemistry, University of California,
Berkeley, CA 94720-1460}

\author{V. Lecomte}

\affiliation{Laboratoire Mati\`ere et Syst\`emes Complexes (CNRS UMR
  7057), Universit\'e de Paris VII, 10 rue Alice Domon et L\'eonie
  Duquet, 75205 Paris cedex 13, France}
			
\author{E. Pitard}

\affiliation{Laboratoire des Collo\"{\i}des, Verres et
Nanomat\'{e}riaux (CNRS UMR 5587), Universit\'e de Montpellier II,
place Eug\`ene Bataillon, 34095 Montpellier cedex 5, France}

\author{K. van Duijvendijk}

\affiliation{Laboratoire Mati\`ere et Syst\`emes Complexes (CNRS UMR
  7057), Universit\'e de Paris VII, 10 rue Alice Domon et L\'eonie
  Duquet, 75205 Paris cedex 13, France}

\author{F. van Wijland}

\affiliation{Laboratoire Mati\`ere et Syst\`emes Complexes (CNRS UMR
  7057), Universit\'e de Paris VII, 10 rue Alice Domon et L\'eonie
  Duquet, 75205 Paris cedex 13, France}

\begin{abstract}
We show that the dynamics of kinetically constrained models of glass
formers takes place at a first-order coexistence line between active
and inactive dynamical phases.  We prove this by computing the
large-deviation functions of suitable space-time observables, such as
the number of configuration changes in a trajectory.  We present
analytic results for dynamic facilitated models in a mean-field
approximation, and numerical results for the Fredrickson-Andersen
model, the East model, and constrained lattice gases, in various
dimensions.  This dynamical first-order transition is generic in
kinetically constrained models, and we expect it to be present in
systems with fully jammed states.
\end{abstract}


\maketitle

\begin{figure}[t]
\begin{centering}
\includegraphics[width=8cm]{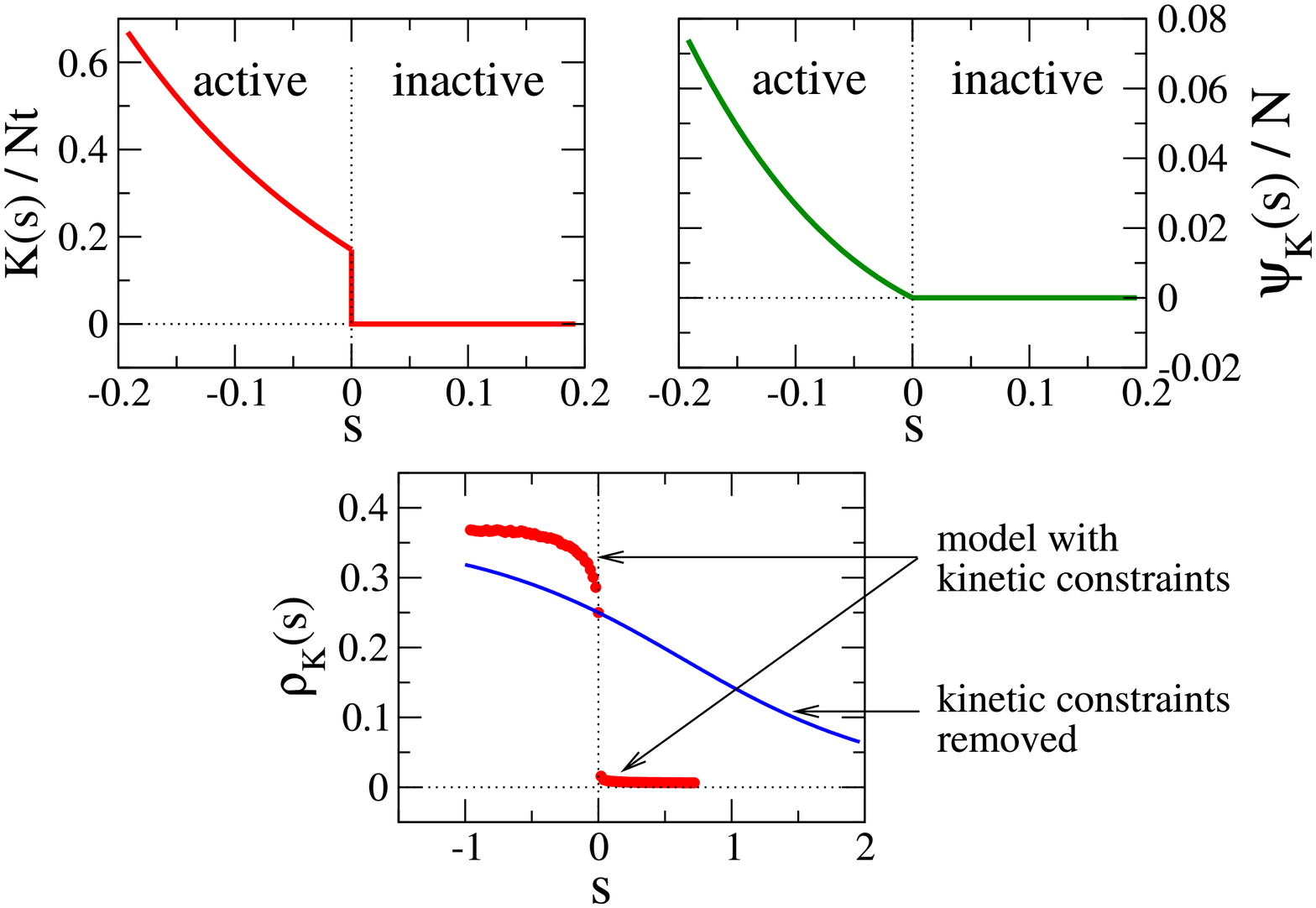}
\caption{First order transition in terms of the field $s$.  (Top) The
dynamical order parameter $K(s)$ (the average number of configuration
changes in a trajectory) and its large-deviation function $\psi_K(s)$
for the FA model, calculated in a mean-field approximation, for $d=3$
and $T=0.5$: see Eqs.\ (\ref{WKFA}-\ref{psiKmf}).  The large deviation
function is singular at $s=0$ and the order parameter $K$ has a
first-order jump.  The dynamics has two phases, an active one for
$s<0$ and an inactive one for $s>0$.  Physical dynamics take place at
$s=0$, where the two dynamic phases coexist.  (Bottom) An alternative
order parameter $\rho_K(s)$ (the average number of excited sites in a
trajectory: see Fig.~\ref{fig2}) 
in the $d=1$ FA model at $T=0.91$, calculated numerically
in a finite system ($N=100$ sites).  The transition is absent when the
kinetic constraints are removed.  }
\label{fig1}
\end{centering}
\end{figure}

An increasingly accepted view is 
that the phenomenology associated with the glass transition \cite{Glass} 
requires a purely dynamic analysis, and 
does not arise from an underlying static transition (see however \cite{Wolynes}).  Indeed, 
it has been suggested that the glass transition manifests a
first-order phase transition in space and time between active and
inactive phases \cite{Merolle-Jack}.  
Here we apply Ruelle's
thermodynamic formalism \cite{Ruelle,Lecomte} to show that this
suggestion is indeed correct, for a specific class of stochastic
models.  The existence of active and inactive regions of
space-time, separated by sharp interfaces, is dynamic heterogeneity,
a central feature of glass forming systems~\cite{DH}.
This phenomenon, in which the dynamics becomes increasingly
spatially correlated at low temperatures, 
arises naturally~\cite{Garrahan-Chandler}  
in models based on the idea of dynamic facilitation,
such as spin-facilitated
models~\cite{Fredrickson-Andersen,Jackle}, constrained lattice
gases~\cite{Kob-Andersen,Jackle-tlg} and other kinetically constrained
models (KCMs)~\cite{Ritort-Sollich}. 
Fig.~\ref{fig1} illustrates the discontinuities in space-time
order parameters at the dynamical transition
in one such model, together with the singularity in a
space-time free energy, as a function of a control parameter to be
discussed shortly.  

The thermodynamic formalism of Ruelle and coworkers was developed in
the context of deterministic dynamical systems \cite{Ruelle}.  While
traditional thermodynamics is used to study fluctuations associated
with configurations of a system, Ruelle's formalism yields information
about its trajectories (or histories).  The formalism relies on the
construction of a dynamical partition function, analogous to the
canonical partition function of thermodynamics.  The energy of the
system is replaced by the dynamical action (the negative of the
logarithm of the probability of a given history); the entropy of the
system by the Kolmogorov-Sinai entropy \cite{Glotzer}, and the
temperature by an intrinsic field conjugate to the action. 
This formalism has been exploited recently to describe 
the chaotic properties of continuous-time Markov processes \cite{Lecomte}.

\begin{figure*}[t]
\begin{centering}
\epsfig{file=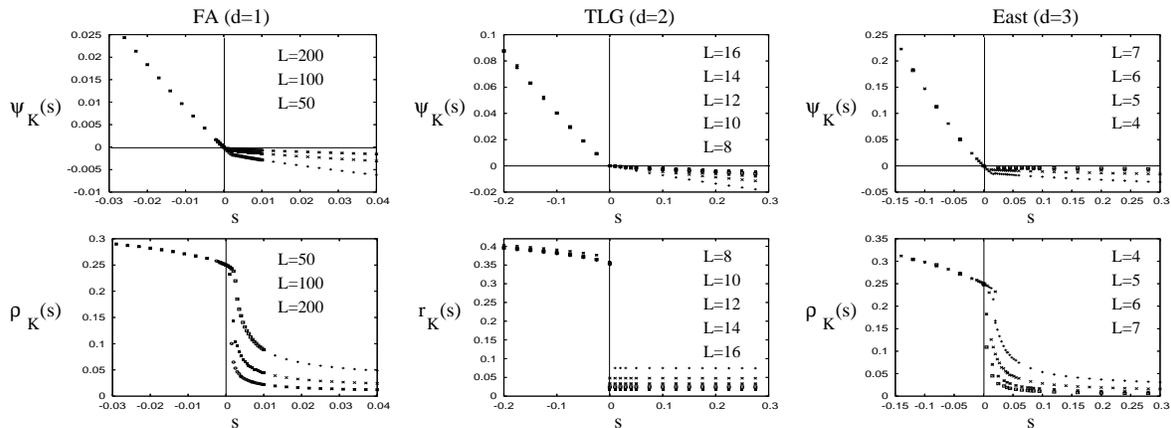,width=1.8\columnwidth}
\caption{
Large-deviation function $\psi_K$ and order
parameters as a function of system size, for the FA model ($d=1)$, the
TLG ($d=2$) and the East model ($d=3$).
For the FA and East models,
we show the average density of excited sites
$\rho_K(s) \equiv Z(s,t)^{-1} \sum_\mathrm{histories}
\mathrm{Prob}(\mathrm{history})e^{-s\hat{K}(\mathrm{history})}
\rho(\mathrm{history})$, where $\rho(\mathrm{history})=
(Nt)^{-1}\int_0^{t}\!\mathrm{d}t' \sum_i n_i(t')$.
For the TLG, we replace $\rho(\mathrm{history})$ by
$r(\mathrm{history})=(Nt)^{-1}\int_0^{t}\!\mathrm{d}t'\, r(\C(t'))$,
to obtain the averaged escape rate $r_K(s)$.
We use representative
conditions: $T=0.91$ for the FA and East models, and density $0.5$ in
the TLG.  Results were obtained by the dynamical method of
\cite{clones}.
}
\label{fig2}
\end{centering}
\end{figure*}

In this work, we define the dynamical partition
sum~\cite{Ruelle,Lecomte} for our stochastic systems by
\begin{equation}
Z_K(s,t) = \sum_{\mathrm{histories}} {\rm Prob}(\mathrm{history})
e^{-s \hat{K}(\mathrm{history})} ,
\label{Z}
\end{equation}
where the sum is over histories from time 0 to time $t$; the
probability of a history is ${\rm Prob}(\mathrm{history})$; and
$\hat{K}(\mathrm{history})$ is the number of configuration changes in
that history. 
$\hat{K}(\mathrm{history})$ is a direct measure of the activity in a history: 
an active trajectory has many changes of configuration, an inactive one has few or none.
At long times
\begin{equation}
Z_K(s,t) \propto e^{t \psi_K(s)} ,
\label{Zst}
\end{equation}
where $\psi_K(s)$ is the large deviation function for $\hat{K}$. The
quantity $-\psi_K(s)$ is a free energy per unit time, for trajectories
\cite{Ritort-Peliti}.  When $t$ is large, the derivatives of
$\psi_K(s)$ give the cumulants of $\hat{K}$.  For example, the mean
number of configuration changes in a trajectory of length $t$ is
$K(0)$, where
\begin{equation}
K(s) = -t \frac{\mathrm{d}}{\mathrm{d}s}\psi_K(s) . 
\end{equation}
For some models, it is reasonable to use the dynamical action, or
dynamical complexity, $\hat{Q}_+ \equiv \ln{[\rm
Prob(\mathrm{history})]}$~\cite{Lecomte}, as an order parameter for
dynamical activity, in place of $\hat{K}$.  In that case the large
deviation function is the topological pressure and the average of
$(-\hat{Q}_+/t)$ at $s=0$ is the Kolmogorov-Sinai entropy
\cite{Ruelle,Lecomte}.  In \cite{Merolle-Jack},
the action was denoted by $\mathcal{E} \equiv -Q_+$ and the
quantity $1+s$ was denoted by $b$.  In the following,
we focus on $\hat{K}$, which simplifies the analysis.

We consider facilitated models \cite{Ritort-Sollich} defined by a
binary field $n_i=0,1$ and a non-interacting Hamiltonian $H = \sum_i
n_i$, where $i=1,\ldots,N$ are the sites of a lattice.  The dynamics
at site $i$ is subject to a kinetic constraint $C_i(n_j)$ which is a
function of the neighbours $n_j$ of $i$.  That is, site $i$ changes
its state with a rate proportional to $C_i$.  For example, in the
Fredrickson-Andersen (FA)
model~\cite{Fredrickson-Andersen,Ritort-Sollich}, $C_i=1$ if any of
the nearest neighbours $j$ of $i$ are in the ``excited'' state,
$n_j=1$; otherwise $C_i=0$.  Sites with $C_i=1$ make the transitions
$0 \rightarrow 1$ with rate $c$ and $1 \rightarrow 0$ with rate
$(1-c)$, where $c$ is the equilibrium concentration of excitations at
temperature $T$, $c \equiv \langle n_i \rangle = (1+e^{1/T})^{-1}$.
One can also consider exchange dynamics $1_i 0_j
\leftrightarrow 0_i 1_j$ between nearest neighbour sites $i$ and $j$,
so that the total density of excited sites is conserved.  The result
is a constrained lattice gas
\cite{Kob-Andersen,Jackle-tlg,Ritort-Sollich}.  For example, the two vacancy facilitated lattice gas, or (2)-TLG \cite{Jackle-tlg}, is defined on a triangular lattice,
with exchange dynamics and a constraint $C_{ij}$ which vanishes unless
the two common nearest neighbours of sites $i$ and $j$ are vacant.

Dynamic heterogeneity in KCMs is a consequence of space-time
correlations in the dynamics.  Within the glassy regime, trajectories
contain space-time regions with relatively fast dynamics, and others
which rearrange slowly \cite{Garrahan-Chandler}.  Slow regions can be
non-finite in extent, that is, KCMs contain states for which the
number of sites with $C_i \neq 0$ is subextensive in the system size
$N$.  These configurations are important for our purposes.
Transitions out of these states occur with a rate that is subextensive
in $N$.  For systems obeying detailed balance, we show below that if
any such states exist the partition sum (\ref{Z}) is dominated by
trajectories localised in them, at large $N$ and $t$ and for all
positive $s$.  The order parameter $\hat{K}$ is subextensive for these
trajectories, so that $K(s)/N$ vanishes for $s>0$ in the large $N$
limit.  Conversely, for $s<0$, the dominant trajectories visit states
in which a finite density of sites have $C_i \neq 0$, and $K(s)/N$ is
finite.  These arguments establish the existence of the transition
shown in Fig.~\ref{fig1}.

For a more quantitative analysis, we identify $\psi_K(s)$ as the
largest eigenvalue of a time evolution operator
\cite{Lebowitz-Spohn,Lecomte}.  The dynamics of the stochastic model
are specified by a master equation
\begin{equation}
\partial_t P(\C,t) = -r(\C) P(\C,t) + \sum_{\C'} W(\C'\to \C) P(\C',t)
,
\end{equation}
where $\C$ denotes a configuration of the system, the $W(\C' \to \C)$
are the transition probabilities between configurations, and
$r(\C)=\sum_{\C'}W(\C\to\C')$ is the rate of escape from $\C$.  We
write this equation in an operator form $\partial_t P = \mathbb{W} P$,
where the matrix elements of $\mathbb{W}$ are
$\mathbb{W}(\C,\C')=W(\C'\to \C)-r(\C)\delta_{\C,\C'}$.

Now, consider the probability $P(\C,t,K)$ that the system is in
configuration $\C$ at time $t$, having made $K$ transitions since time
$0$.  We define $\tilde{P}(\C,t,s) \equiv
\sum_K e^{-s K} P(\C,t,K)$, so that $Z(s,t)=\sum_{\C}
\tilde{P}(\C,t,s)$.  The time evolution of $\tilde{P}$ obeys a Master
equation $\partial_t \tilde{P} = \mathbb{W}_K \tilde{P}$, where
\cite{Lebowitz-Spohn,Lecomte}
\begin{equation}
\mathbb{W}_K(\C,\C') = e^{-s} W(\C' \to \C) -
r(\C) \delta_{\C,\C'} .
\label{WKs}
\end{equation} 
The calculation of $\psi_K(s)$ reduces to finding the largest eigenvalue
of $\mathbb{W}_K$.

The FA model dynamics can be described using the Doi-Peliti
\cite{Doi-Peliti} occupation number formalism.  The operator
(\ref{WKs}) then reads \cite{Whitelam,Jack-Mayer}:
\begin{equation}
\mathbb{W}^{{\rm (FA)}}_K = \sum_{\langle i,j \rangle} \hat{n}_i
\left[ c \left( e^{-s} a_j^\dagger -1 \right) + \left( e^{-s} -
a_j^\dagger \right) a_j \right] .
\label{WKFA}
\end{equation}
where $a_i^\dagger$ and $a_i$ are bosonic creation and annihilation
operators, $\hat{n}_i = a_i^\dagger a_i$ and $\langle i,j \rangle$
indicates that the sum is over pairs of nearest neighbours.  For
simplicity we have allowed for multiple occupancy per site: changing
the dynamics in this way has little effect at low temperatures
\cite{Whitelam}.  The single occupancy case can be formulated in a
similar manner \cite{Jack-Mayer}, and the results are analogous.

We begin with a simple mean-field treatment of $\mathbb{W}^{{\rm
(FA)}}_K$.  We discard fluctuations and define
$W^{(\mathrm{FA})}_K(\phi,\phibar)$ by making the replacements $a_i
\to \langle a_i \rangle \equiv \phi$ and $a_i^\dagger \to \langle
a_i^\dagger \rangle \equiv \phibar$ in the operator
$\mathbb{W}^{(\mathrm{FA})}_K$.  Then we can estimate the largest
eigenvalue of $\mathbb{W}^{{\rm (FA)}}_K$ by finding the stationary
points of $W^{(\mathrm{FA})}_K(\phi,\phibar)$.  The Euler-Lagrange
equations are
\begin{eqnarray}
\label{equ:EL1}
0 &=& 2 \phibar \phi \left( \phi - e^{-s} c \right) + \phi \left( c -
      e^{-s} \phi \right) , \\ 0 &=& 2 \left( \phibar - e^{-s} \right)
      \phibar \phi + c \phibar \left( 1 - e^{-s} \phibar \right) .
\label{equ:EL2}
\end{eqnarray}
There are two solutions that correspond to steady states of the
dynamics: (i) $\phi=c \phibar=c\phibar_*$ with $\phibar_* =
\frac{3}{4} e^{-s} + \frac{1}{4} \sqrt{9 e^{-2s} - 8}$; and (ii)
$\phibar=\phi=0$.  For solution (i),
$W^{(\mathrm{FA})}_K(\phi,\phibar)$ is positive for $s<0$ and negative
otherwise; for solution (ii), $W^{(\mathrm{FA})}_K(\phi,\phibar)=0$.
Our estimate for the large deviation function is given by the larger
value of $W^{(\mathrm{FA})}_K(\phi,\phibar)$, so that
\begin{equation}
\psi_K^{(\rm m.f.)}(s) = \left \{ \begin{array}{cc} Nd \left( c
\phibar_* \right)^2 \left(e^{-s} \phibar_* - 1 \right) & (s<0) \\ 0 &
(s>0) \\
\end{array} \right.
\label{psiKmf}
\end{equation}
where $d$ is the spatial dimension, and $\phibar_*$ is defined above.
Figure \ref{fig1}(top) shows $\psi_K^{(\rm m.f.)}(s)$ and $K(s)$.  The
active phase is stable for $s<0$; the inactive one is stable for
$s>0$.  The large deviation function $\psi_K^{(\rm m.f.)}(s)$ is
continuous at $s=0$, but its derivative $K(s)$ displays a first-order
jump.

The mean-field approximation is clearly very crude.  It can be
improved systematically with loop corrections, but the coexistence
scenario does not change qualitatively: it is a consequence of the
kinetic constraints and the existence of states with subextensive
escape rates.  We now confirm this by means of numerical simulations.

The time evolution associated with the operator $\mathbb{W}_K$ can be
obtained dynamically. This operator does not conserve probability [see
Eq.\ (\ref{WKs})] but there are methods to simulate the evolution that
it represents \cite{Giardina}.  In our case we
apply this scheme, analogous to the quantum diffusion Monte Carlo procedure, 
to continuous time Monte Carlo dynamics 
\cite{clones}. In Fig.\ \ref{fig1} we show a plot of $\rho_K(s)$
for the FA model in one dimension.  In Fig.\
\ref{fig2} we present a finite size scaling analysis of both $\psi_K$
and a dynamical order parameter for the FA model in one dimension, the
(2)-TLG model in two dimensions, and the East model in three dimensions (see e.g.\ \cite{Berthier-Garrahan}).  As we
approach $N \to \infty$ the crossover between
active and inactive phases becomes sharp.  Figure~\ref{fig2} shows
results for three different models, in different dimensions and with
different dynamical constraints; the TLG has a conserved density while
the FA and East models do not.  Our similar results for these
different models demonstrate that space-time phase coexistence occurs
quite generally in KCMs.

\begin{figure}[t]
\begin{centering}
\includegraphics[width=\columnwidth]{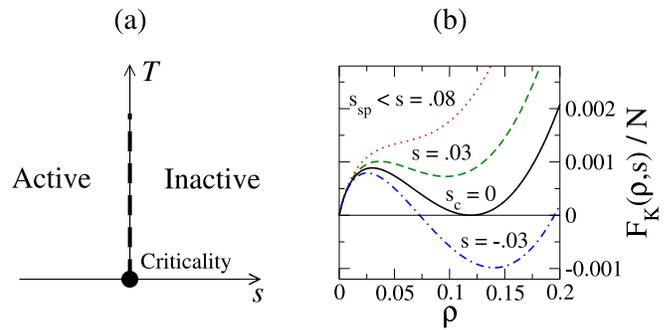}
\caption{(a) Space-time phase diagram for facilitated models.  $s=0$
  is a first-order transition line between the active phase ($s<0$)
  and the inactive phase ($s>0$).  It terminates in a critical point
  at $T=0$.  (b) Mean-field variational free-energy
  $\mathcal{F}_K(\rho,s)$, Eq.\ (\ref{FKs}), for the FA model at
  $T=0.5$, for several values of $s$.  For $s<0$ the active phase is
  dominant and the absolute minimum of $\mathcal{F}_K$ is at $\rho
  \neq 0$.  At $s=0$ there is dynamic phase-coexistence.  For $s>0$
  the inactive phase dominates and $\rho=0$ minimizes the free-energy.
  The active minimum no longer exists beyond the spinodal $s_{\rm
  sp}$.}
\label{fig3}
\end{centering}
\end{figure}

We have demonstrated that, for any temperature $T$, the dynamics of KCMs 
such as the FA or East models take place at dynamical phase coexistence
between active and inactive phases.  We can construct a phase diagram
in terms of $T$ and $s$, see Fig.\ \ref{fig3}(a).  The $s=0$ axis is a
first-order transition line.  It ends in a critical point at $T=0$.
For constrained lattice gases such as the Kob-Andersen model and the
(2)-TLG, the phase diagram is similar, with $T$ replaced by density of
particles $c$.  In that case the critical point is at the maximum
density $c=1$.

We emphasize that while a zero temperature dynamical critical point is
common to many KCMs, this is not a sufficient condition for space-time
phase coexistence.  For example, consider the pair appearence and annihilation (AA) model of
\cite{Jack-Mayer}, which has the same critical properties as the FA
model.  Its dynamical rules are $\emptyset A
\rightleftharpoons A \emptyset$, $AA
\rightarrow \emptyset$, $\emptyset \emptyset
\rightarrow AA$, with rates $D$, $\lambda$ and $\gamma$, respectively.  All states in this model
have extensive escape rates, so we do not expect any transition
at $s=0$.  Deriving the
Euler-Lagrange equations analogous to (\ref{equ:EL1},\ref{equ:EL2}),
leads to a large deviation function that is analytic at $s=0$.  We
have also calculated $\psi_K(s)$ exactly \cite{longPaper} for the AA
model in $d=1$, at the free fermion point
$\lambda+\gamma=2D$~\cite{freeF}: the large deviation function
$\psi_K(s)$ is indeed analytic at $s=0$~\cite{longPaper}.

Finally, in models that obey detailed balance with respect to a
probability distribution $p_\C$, such as the ones considered here,
$\psi_K(s)$ can be calculated through a variational method.  The
master operator Eq.\ (\ref{WKs}) is made symmetric by a similarity
transformation, $\mathbb{H}_K(\C,\C') \equiv p^{-1/2}_{\C}
\mathbb{W}_K(\C,\C') p^{1/2}_{\C'} = e^{-s} \sqrt{W(\C' \to \C) W(\C
\to \C')} - r(\C) \delta_{\C,\C'}$.  Since $\mathbb{H}_K$ is symmetric
and has the same eigenvalues as $\mathbb{W}_K$, we can apply a
variational principle:
\begin{equation}
\psi_K(s)= \max_{V_\C} \ \frac{\sum_\C V_\C \mathbb{H}_K(\C,\C')
V_{\C'}}{\sum_\C V_\C V_\C} .
\label{equ:var}
\end{equation}

Using a trial distribution in which only one of the $V_\C$ is finite
shows that the largest diagonal element of $\mathbb H$ is a lower
bound on $\psi_K(s)$, so that $\psi_K(s) \geq -\min_\C r(\C)$.  Hence, if
there exists a state for which the escape rate $r(\C)$ is
subextensive, then $\psi_K(s)/N \geq 0$ at large $N$.  Further,
$\psi_K(s)$ is non-increasing [since $\hat{K}$ is non-negative] and
$\psi_K(0)=0$, so the existence of a subextensive escape rate
establishes immediately that $\psi_K(s)/N=K(s)/N=0$ for all $s>0$ in the
large $N$ limit, as we asserted above.

As an example of how to obtain $\psi_K(s)$ from Eq.\ (\ref{equ:var})
consider the FA model in a mean-field geometry, such as the complete
graph.  The transition rates in this case are $W(n \to n+1)=cn$,
$W(n+1 \to n)=n (n+1)/N$, where $n=\sum_i n_i$ is the total number of
excitations.  For the variational state we assume $V_n = e^{N
f(n/N)}$, for some function $f(\rho)$ of the excitation density $\rho
\equiv n/N$.  In the limit of large $N$, the leading contribution to
(\ref{equ:var}) comes from values of $\rho$ that maximize $f(\rho)$,
and Eq.\ (\ref{equ:var}) reduces to $\psi_K(s) = - \min_{\rho}
\mathcal{F}_K(\rho,s)$, where
\begin{equation}
\mathcal{F}_K(\rho,s) = N \rho \left( \rho - 2 e^{-s} \sqrt{c \rho} +
c  \right) .
\label{FKs}
\end{equation}
$\mathcal{F}_K$ is a Landau free-energy function for the order
parameter $\rho$.  The minimum of $\mathcal{F}_K$ occurs at $\rho_*=c
(\phibar_*)^2$, and from Eq.\ (\ref{equ:var}) the result of Eq.\
(\ref{psiKmf}) is recovered.  As shown in Fig.\ \ref{fig3}(b) the
function $\mathcal{F}_K(\rho,s)$ behaves in the characteristic way
associated with a first-order phase-transition.  

We have shown that the dynamics of KCMs is characterised by the
coexistence in space-time of active and inactive dynamical phases.  
In our view, this dynamical phase coexistence underlies the heterogeneous 
particle dynamics observed in glass formers.  Thus, 
experimentally
obervable phenomena 
such as transport decoupling~\cite{DH}
arise from the fluctuations associated with this
dynamic phase equilibrium.  The extension~\cite{Merolle-Jack} of our
results to atomistic models will clarify the
degree to which the theoretical framework described here captures 
the generic features of glassy dynamics.

\smallskip

We are indebted to David Chandler for encouragement and for many
suggestions on the manuscript.  This work was supported by EPSRC
grants no.\ GR/R83712/01 and GR/S54074/01 (JPG); NSF grant CHE-0543158
(RLJ); and a French Ministry of Education ANR grant JCJC-CHEF.

\end{document}